\documentclass[aps,pre,twocolumn,superscriptaddress,amsmath,showpacs,amssymb]{revtex4}
\usepackage{epsfig}

\def\cal#1{\mathcal{#1}}
\newcommand{\av}[1]{\left\langle #1\right\rangle}
\newcommand{\bra}[1]{\left\langle #1\right|}
\newcommand{\ket}[1]{\left|#1\right\rangle}

\newcommand{\bracket}[3]{\left\langle #1\left|#2\right|#3\right\rangle}
\def\ad{a^{\dagger}}
\def\n0{\bar{n}_0}
\def\ps{\phi^{\star}}
\def\pb{\bar{\phi}}

\def\goo{\Gamma^{(1,1)}}
\def\got{\Gamma^{(1,2)}}
\def\gto{\Gamma^{(2,1)}}
\def\gtt{\Gamma^{(2,2)}}
\def\beq{\begin{equation}}
\def\eeq{\end{equation}}
\def\bea{\begin{eqnarray}}
\def\eea{\end{eqnarray}}

\begin{document}

\title{Renormalization group study of a kinetically constrained model
for strong glasses}

\author{Stephen Whitelam} 

\affiliation{Rudolf Peierls Centre for Theoretical Physics, University
of Oxford, 1 Keble Road, Oxford, OX1 3NP, UK}

\author{Ludovic Berthier} 

\affiliation{Rudolf Peierls Centre for Theoretical Physics, University
of Oxford, 1 Keble Road, Oxford, OX1 3NP, UK}

\affiliation{Laboratoire des Verres UMR 5587, Universit\'e Montpellier
II and CNRS, 34095 Montpellier, France}

\author{Juan P. Garrahan}

\affiliation{School of Physics and Astronomy, University of
Nottingham, Nottingham, NG7 2RD, UK}

\date{\today}

\begin{abstract}
We derive a dynamic field theory for a kinetically constrained model,
based on the Fredrickson--Andersen model, which we expect to describe
the properties of an Arrhenius (strong) supercooled liquid at the
coarse-grained level. We study this field theory using the
renormalization group. For mesoscopic length and time scales, and for
space dimension $d \geq 2$, the behaviour of the model is
governed by a zero-temperature dynamical critical point in the
directed percolation universality class. We argue that in $d=1$ its
behaviour is that of compact directed percolation.  We perform
detailed numerical simulations of the corresponding
Fredrickson-Andersen model on the lattice in various dimensions, and
find reasonable quantitative agreement with the field theory
predictions.
\end{abstract}

\maketitle

\section{Introduction}

The discovery that supercooled liquids, whose structures are
essentially homogeneous and featureless, are dynamically highly
heterogeneous is arguably the most important recent development in the
long-standing problem of the glass
transition~\cite{Ediger-et-al,Angell,Debenedetti-Stillinger}.  Dynamic
heterogeneity has been observed both experimentally in deeply
supercooled liquids~\cite{DHexp1}, and in numerical simulations of
mildly supercooled liquids~\cite{DHnum,Vogel-Glotzer}.  In addition,
dynamic heterogeneity has been observed experimentally in colloidal
suspensions~\cite{Weeks-et-al}.  For recent reviews
see~\cite{Sillescu,Ediger,Glotzer,Richert}.
 
Understanding dynamic heterogeneity is a crucial step towards
understanding the glass transition. Dynamic heterogeneity implies that
the slow dynamics of glass-formers is dominated by spatial
fluctuations, a feature discarded from the start in homogeneous
approaches like mode coupling theories \cite{MCT} or other mean-field
treatments~\cite{Mezard-Parisi}; see
however~\cite{Franz-Parisi,Biroli-Bouchaud,Szamel,Schweizer-Saltzman}.
Moreover, the absence of growing structural length scales has been the
main obstacle to the application to supercooled liquids of many of the
tools used so successfully to analyze conventional phase
transitions. The existence of dynamic heterogeneity, however, implies
that the increase in timescales as the glass transition is approached
is associated with growing length scales of dynamically, not
statically, correlated regions of space~\cite{Garrahan-Chandler}
(Refs.~\cite{Tarjus-Kivelson,Xia-Wolynes} offer alternative
thermodynamic viewpoints). This suggests that supercooled liquids
might display universal dynamical scaling, by analogy with
conventional dynamical critical phenomena. This scaling behaviour
could then be studied by standard renormalization group (RG)
techniques.  This is the issue we address in detail in this paper,
which is a follow-up to our recent Letter~\cite{Whitelam-et-al}.

Our starting point is the coarse-grained real-space description of
glass-formers developed in
\cite{Garrahan-Chandler,Garrahan-Chandler-2,Berthier-Garrahan} which
places dynamic heterogeneity at its core. It is a mesoscopic approach,
based on two observations. First, at low temperature very few
particles in a supercooled liquid are mobile, and these mobility
excitations are localized in space; and second, mobile regions of the
liquid are needed to allow neighbouring regions to themselves become
mobile. This second observation is the concept of dynamic
facilitation~\cite{Glarum,Fredrickson-Andersen,Palmer-et-al,Ritort-Sollich}. This
picture of supercooled liquids can be straightforwardly cast as a
dynamical field theory and its scaling behaviour determined via RG. We
find that, generically, scaling properties are governed by a
zero-temperature critical point. We study in detail the simplest case,
that of isotropic dynamic facilitation, which we expect to model an
Arrhenius, or strong, glass-former. We show that its low temperature
dynamics is controlled by a zero-temperature critical point, which for
dimensions $d \geq 2$ is that of directed percolation
(DP)~\cite{Hinrichsen}. We argue that in $d=1$ the model belongs to
the universality class of compact directed percolation
(CDP)~\cite{Hinrichsen}.  Our theoretical predictions compare
favourably with our results from numerical simulations of the
Fredrickson-Andersen (FA) model~\cite{Fredrickson-Andersen}, the
lattice model on which the field theory is based.

This paper is organized as follows. We derive a field theoretic
description of a generic system possessing constrained dynamics in
Section II, and we discuss in Section III the physical interpretation
of the field theory for the special case of an
isotropically-constrained model. In Section IV we study this field
theory for $d \geq 2$ using RG, and in Section V we discuss the
special case of $d=1$. In Section VI we compare the theoretical
predictions to simulations of the FA model in various
dimensions. Section VII contains a summary of our results and
conclusions.

\section{Derivation of the field theory} 

We build an effective model for glass-formers as
follows~\cite{Garrahan-Chandler-2}. We coarse-grain a supercooled
fluid in $d$ spatial dimensions into cells of linear size of the order
of the static correlation length, as given by the pair correlation
function. We assign to each cell a scalar mobility, $n_i$, whose value
is chosen by further coarse-graining the system over a microscopic
time scale. Mobile regions carry a free energy cost, and when mobility
is low we do not expect interactions between cells to be
important. Adopting a thermal language, we expect static equilibrium
to be determined by a non-interacting
Hamiltonian~\cite{Ritort-Sollich},
\begin{equation}
\label{hamiltonian}
H = \sum_{i=1}^N n_i .
\end{equation}
At low mobility, the distinction between single and multiple occupancy
is probably unimportant, and we assume the latter case for technical
simplicity. The question of whether the field theories for versions of
a system with single (`fermionic') or multiple (`bosonic') occupancies
lie in the same universality class is unresolved~\cite{Brunel-et-al},
but if the distinction matters it is likely to matter more in $d=1$
than in higher dimensions. We shall ignore this subtlety.

We define the dynamics of the mobility field by a master equation,
\begin{equation}
\label{master1}
\partial_t P\left( n  , t \right) =
\sum_i \mathcal{C}_i \left(  n  \right) \,
\hat{\mathcal{L}_i} \, P\left(  n  , t \right) ,
\end{equation}
where $P\left( n , t \right)$ is the probability that the system has
configuration $n \equiv \{n_i\}$ at time $t$. Equation (\ref{master1})
shows clearly the two ingredients of our model. The first, the
existence of local quanta of mobility, is encoded by the local
operators $\hat{\mathcal{L}_i}$. For non-conserved dynamics we choose
these to describe creation and destruction of mobility at site $i$,
\begin{eqnarray}
\label{famaster}
\hat{\mathcal{L}_i} \, P\left( n_i , t \right) &=& \gamma \, (n_i+1)
P\left( n_i+1, t \right) + \rho \, P\left( n_i-1, t \right) \nonumber
\\ && -(\gamma \, n_i + \rho) P\left(n_i, t \right) ,
\end{eqnarray} 
where the dependence of $P$ on cells other than $i$ has been
suppressed. The rates for mobility destruction, $\gamma$, and
creation, $\rho$, are chosen so that (\ref{master1}) obeys detailed
balance with respect to (\ref{hamiltonian}) at low temperature. This
means that the stationary solution of the master equation must equal
the Gibbs distribution. Equation (\ref{hamiltonian}) gives rise to the
Gibbs distribution $P_{\rm eq}(n)=\prod_i (1-e^{-1/T}) e^{-n_i/T}$,
whereas the master equation (\ref{master1}) has the stationary
solution
\begin{equation}
\label{stat}
P(n,t \rightarrow \infty)= \prod_i e^{-\rho/\gamma} \left(
\frac{\rho}{\gamma} \right)^{n_i} \frac{1}{n_i!}.
\end{equation} 
Equation (\ref{stat}) will reduce to $P_{\rm eq}(n)$ provided
$\rho/\gamma= e^{-1/T}$ and $T \ll 1$. Thus detailed balance with
respect to (\ref{hamiltonian}) holds only at low temperature, and we
will hereafter assume that $T \ll 1$. For convenience we write $
\rho/\gamma \approx c$, where $c \equiv \langle n_i \rangle$;
angle brackets $\av{\cdots}$ denote an equilibrium, or thermal,
average. The thermal concentration of excitations $c$ is the control
parameter of the model.

The second ingredient of our model is the kinetic constraint,
$\mathcal{C}_i \left( \left\{ n \right\} \right)$, which must suppress
the dynamics of cell $i$ if surrounded by immobile regions. It cannot
depend on $n_i$ itself if (\ref{master1}) is to satisfy detailed
balance. To reflect the local nature of dynamic facilitation we allow
$\mathcal{C}_i$ to depend only on the nearest neighbours of
$i$~\cite{Ritort-Sollich} and require that $\mathcal{C}_i$ is small
when local mobility is scarce.

One can derive the large time and length scale behaviour of the model
defined by Eqs.\ (\ref{hamiltonian})--(\ref{famaster}) from an
analysis of the corresponding field theory. The technique to recast
the master equation (\ref{famaster}) as a field theory is
standard~\cite{Doi-Peliti,Cardy-et-al}. One introduces a set of
bosonic creation and annihilation operators for each site $i$,
$a^{\dagger}_i$ and $a_i$, satisfying
$[a^{\dagger}_i,a_j]=\delta_{ij}$, and defines a set of states
$\ket{n}= (\ad)^n \ket{0}$, such that
\begin{equation}
\ad_i \ket{n_i}=\ket{n_i+1}, \qquad a_i \ket{n_i} = n_i \ket{n_i-1}.
\end{equation}
The vacuum ket $\ket{0}$ is defined by $a \ket{0}=0$. One passes to a
Fock space via a state vector
\begin{equation}
\label{state-vec}
|\Psi(t) \rangle \equiv \sum_{\{n_i\}}
P(n,t) \prod_i a_i^{\dagger \,n_i} \ket{0}.
\end{equation}
The master equation (\ref{famaster}) then assumes the form
of a Euclidean Schr\"{o}dinger equation, 
\begin{equation}
\partial_t |\Psi(t) \rangle =-\hat{H} |\Psi(t) \rangle,
\end{equation}
with $\hat{H}=\sum_i
\hat{\mathcal{C}}_i(\{a^{\dagger}_j a_j\}) \hat{H}_i^{(0)}$. The
unconstrained piece $\hat{H}^{(0)}_i$ reads
\begin{equation}
\label{noncon}
\hat{H}^{(0)}_{i}=-\gamma (a_i-a^{\dagger}_i a_i)-\rho
(a^{\dagger}_i-1) ,
\end{equation}
which describes the creation and destruction of bosonic excitations
with rates $\rho$ and $\gamma$. The evolution operator $e^{-\hat{H}
t}$ can then be represented as a coherent state path
integral weighted by the dynamical action~\cite{Cardy-et-al}
\begin{equation}
\label{action1}
\mathcal{S}[\phi^{\star}_i,\phi_i,t_0] = \sum_i \int_0^{t_0} dt \,
{\big\lgroup} \phi^{\star}_i \partial_t \phi_i +
H_i(\phi^{\star}, \phi) {\big\rgroup} ,
\end{equation} 
where we have suppressed boundary terms coming from the system's
initial state vector. The fields $\phi^{\star}_i(t)$ and $\phi_i(t)$
are the complex surrogates of $a^{\dagger}_i$ and $a_i$, respectively,
but must now be treated as independent fields and not complex
conjugates. The Hamiltonian $H_i$ has the same functional form as
(\ref{noncon}) with the bosonic operators replaced by the complex
fields. At the level of the first moment we have $\langle n_i \rangle
= \langle \phi_i \rangle$, and so we may regard $\phi_i$ as a complex
mobility field. Higher moments of $n_i$ and $\phi_i$ are not so simply
related, however: for example, $\langle n_i^2 \rangle = \langle
\phi_i^2 \rangle + \langle \phi_i \rangle$~\cite{Mattis-Glasser}. The
last step in the passage to a field theory is to take the continuum
limit, according to $\sum_i \rightarrow a^{-d} \int d^{d}x$, $
\phi_i(t) \rightarrow a^d \phi(\mathbf{x},t)$, and $ \phi^{\star}_i(t)
\rightarrow \phi^{\star}(\mathbf{x},t)$, where $a$ is the lattice
parameter.

The definition of the model is completed by specifying the functional
form of the kinetic constraint. The simplest non-trivial form is an
isotropic facilitation function, $\mathcal{C}_i = \sum_j n_j$, where
the sum is over nearest neighbours of site $i$. With this choice we
expect our model to be in the same universality class as the one-spin
facilitated Fredrickson-Andersen model in $d$
dimensions~\cite{Fredrickson-Andersen,Ritort-Sollich}. Different
choices for the operators $\hat{H}^{(0)}$ and $\mathcal{C}$ lead to
field theoretical versions of more complicated facilitated models. A
diffusive $\hat{H}^{(0)}$, for example, would correspond to a
constrained lattice gas like that of Kob and Andersen
\cite{Kob-Andersen,Ritort-Sollich}; an asymmetric $\mathcal{C}$ to the
East model \cite{Jackle,Ritort-Sollich} and its generalizations
\cite{Garrahan-Chandler-2}.

In the continuum limit the isotropic constraint reads
\begin{equation}
\label{continuum-c}
\sum_j \ps_j \phi_j \approx \left(2d + a^2 \nabla^2 +\cdots\right )
\ps_i \phi_i,
\end{equation}
where `$\cdots$' denotes higher-order gradient terms irrelevant in the
RG sense in the long time and wavelength limit. Terms linear in the
spatial gradient vanish because the constraint is
isotropic. Consequently, the dynamics of the model is nearly
diffusive, perturbed by fluctuations in low dimensions.

To derive the dynamic action it is convenient to make a linear shift
of the response field, $\ps \to 1+ \pb$~\cite{Cardy-et-al}, in the
Hamiltonian. This is done for the following reason. Expectation values
in this formalism are given by by $\av{A}= \bracket{s}{ A}{\Psi(t)}$,
where $\ket{\Psi(t)}=e^{-\hat{H}t} \ket{\Psi(0)}$ and $\bra{s} \equiv
\bra{0} e^{\sum_i a_i}$~\cite{Mattis-Glasser,Cardy-et-al}. The 
projection state $\bra{s}$ is introduced because the usual quantum
mechanical expression $\bracket{\Psi(t)}{A}{\Psi(t)}$ is bilinear in
the probability $P$. If one wishes to apply Wick's theorem, one must
commute the factor $e^{\sum_i a_i}$ to the right hand side of the
bracket; the consequent shift $\ps \rightarrow 1 +\pb$ in the
Hamiltonian follows from the identity $e^a f(\ad,a)=f(1+\ad,a) e^a$,
and corresponds to a change of integration variables. It therefore
does not change the properties of the system under
renormalization. However, it can obscure important symmetries of the
model in question, and so should be made with care~\cite{Cardy-et-al}.

The dynamic action now follows from Equations (\ref{noncon}),
(\ref{action1}) and (\ref{continuum-c}), suitably shifted, and reads
\begin{eqnarray}
\label{shift}
\lefteqn{ \mathcal{S}[\bar{\phi},\phi,t_0] = \int d^d x \,
\int_0^{t_0} dt {\big\lgroup} \bar{\phi} \left( \partial_t -D_0
\nabla^2 -r_0 \right)\phi } \nonumber \\ && +\bar{\phi} \phi
(\lambda_0^{(1)}+\nu_0^{(1)} \nabla^2) \phi+\bar{\phi} \phi
(\lambda_0^{(2)}+\nu_0^{(2)} \nabla^2) \bar{\phi} \phi \nonumber \\ &&
-\bar{\phi} \phi (v_0 +\sigma_0 \nabla^2) \bar{\phi} {\big\rgroup}.
\end{eqnarray} 
We have defined $\lambda_0^{(1,2)} \equiv 2 d a^d \gamma$, $r_0=v_0
\equiv 2d \rho$, $\nu_0^{(1,2)} \equiv \gamma a^{d+2}$ and $\sigma_0
\equiv a^{2} \rho$. We write $D_0 \equiv \sigma_0$ to emphasise the
emergence of a diffusive term, although in the unshifted model there
is no purely diffusive process. We have omitted higher-order gradient
terms, and suppressed boundary contributions coming from initial and
projection states. Equation (\ref{shift}) is the starting point for
our RG analysis.

\section{Physical interpretation of the action}

\begin{figure}
\begin{centering}
\psfig{file=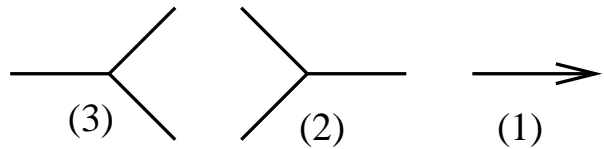,height=8.cm,angle=270}
\caption{\label{fig:elements} Elements of the simplified action
(\ref{simp-action}). From left to right: branching and coagulation
vertices, and diffusive propagator. Time runs from left to
right. These processes appear in the space-time trajectories of
simulations of the lattice FA model in Fig.~\ref{fig:traj}.}
\end{centering}
\end{figure}

Equation (\ref{shift}) has the form of an action for a single species
branching and coagulation diffusion-limited
reaction~\cite{Tauber-review,Hinrichsen} with additional
momentum-dependent terms. We can see how this action governs the
behaviour of the model by dropping all but the most relevant terms
from the action to give (see Section~\ref{sec:rg})
\begin{eqnarray}
\label{simp-action}
\mathcal{S}&=&\int d^dx dt \, \bar{\phi} \left(\partial_t-D_0
\nabla^2-r_0 \right) \phi \nonumber \\ &+& \int d^dx dt \, \left(
\lambda_0 \bar{\phi} \phi^2 - v_0 \bar{\phi}^2 \phi \right).
\end{eqnarray}
The first term is the bare propagator of the theory, the
renormalized version of which corresponds to the probability
that two sites separated in space and time are connected by an
unbroken chain of mobile sites~\cite{Hinrichsen}. The second and
third terms are the vertices corresponding to coagulation and
branching interactions, respectively. In the usual way~\cite{Amit}
one associates with each term in the action a diagram, as in
Fig.~\ref{fig:elements}.

\begin{figure}
\begin{centering}
\psfig{file=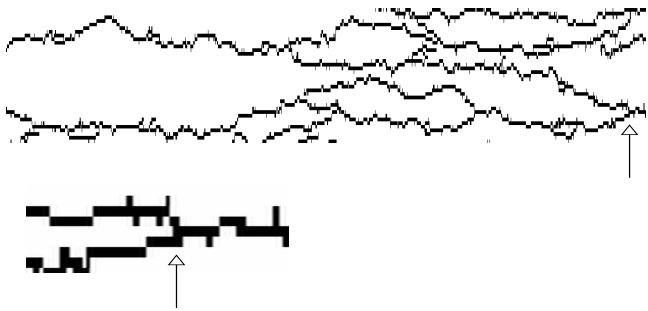,width=8.5cm,height=4.cm}
\caption{\label{fig:traj} A space-time trajectory for the FA model in
$1+1$ dimensions; time runs horizontally from left to right; space
vertically. Mobile sites are black. The events corresponding to the
diagrams in Fig.~\ref{fig:elements} can be clearly seen. The wandering
of the excitation lines corresponds to diffusion; branching and
coagulation events act to renormalize the excitation lines. In the
lower left a coagulation event is shown enlarged.}
\end{centering}
\end{figure}

The physical processes corresponding to the terms in
(\ref{simp-action}) or the diagrams in Fig.~\ref{fig:elements} can be
seen in numerical simulations. In Fig.~\ref{fig:traj} we show a
typical space-time trajectory~\cite{Garrahan-Chandler} for the FA
model in $1+1$ dimensions. The wandering of excitation lines
corresponds to the diffusion of isolated defects. Diffusion appears in
the propagator as a result of the shift $\ps \rightarrow 1+ \pb$
applied to the term $\ps \nabla^2 \ps \phi$, which enters
Eq.~(\ref{action1}). This term corresponds to nearest-neighbour
facilitated mobility creation with rate $\propto c$, and so diffusion
in our model results from facilitated creation (branching), followed
by facilitated destruction (coagulation): $\{ \uparrow \emptyset
\stackrel{c}{\longrightarrow} \uparrow \uparrow
\stackrel{1}{\longrightarrow} \emptyset \uparrow \} \sim c \pb
\nabla^2 \phi$.

Branching and coagulation events can be clearly identified: one of the
latter is enlarged in the lower left of the figure. These events
correspond to fluctuations. In low dimensions, where fluctuations are
important, branching and coagulation events renormalize the bare
propagator of the theory, meaning that excitation lines joining two
sites are dressed by bubbles. In low dimensions one must therefore
resort to RG in order to account for fluctuation effects in a
controlled way.

\section{RG analysis of the action}
\label{sec:rg}

\subsection{Langevin equation of motion}

By making stationary variations of the action (\ref{shift}) with
respect to the response field $\pb$, $\delta \cal{S}/\delta \pb=0$, we
obtain the Langevin equation of motion for the field $\phi$:
\begin{equation}
\label{lang-eqn}
\partial_t \phi(t) = D_0 \nabla^2 \phi+r_0 \phi - \lambda_0 \phi^2 +
\eta(x,t),
\end{equation}
where the noise $\eta$ satisfies $\av{\eta(x,t)}=0$, and 
\begin{equation}
\label{nn-corr}
\av{\eta(x,t) \eta(x',t')} = 2 \left(v_0 \phi-\lambda_0^{(2)}
\phi^2\right)\delta(x-x') \delta(t-t').
\end{equation}
We neglect diffusive noise. The noise-noise correlator (\ref{nn-corr})
describes stochastic fluctuations of the mobility field $\phi$, and
comes from the coefficient of the terms in the action quadratic in
$\pb$. We see that (\ref{nn-corr}) describes a competition between
mobility correlations and anti-correlations, induced by branching and
coagulation, respectively. If for example a branching event occurs, a
particle will find itself with more nearest neighbours than one would
expect from a mean-field argument. In the long time and wavelength
limit, for $d \geq 2$, we show below that the second term in
Eq.~(\ref{nn-corr}) is irrelevant, and may be dropped. Equations
(\ref{lang-eqn}) and (\ref{nn-corr}) then constitute the well-known
Langevin equation for DP~\cite{Hinrichsen}, albeit with a positive
definite mass term.

The mean-field approximation consists of dropping the noise and
diffusion terms from (\ref{lang-eqn}). The resulting equation
possesses a dynamic critical point at $c \propto r_0=0$, or $T=0$. For
$T>0$, in the non-equilibrium regime, the density $\phi$ approaches
its thermal expectation value $c=\rho/\gamma$ exponentially
quickly. At $T=0$ the decay becomes algebraic, $\phi(t \to \infty)
\sim \left(\lambda_0 t \right)^{-1}$. Whether at equilibrium or not,
the mean-field equation admits the critical exponents
$\nu_{\parallel}=1$ and $\beta=1$. The former describes the growth of
time scales $\xi_{\parallel}$ near criticality, via
$\xi_{\parallel}\sim c^{-\nu_{\parallel}}$; the latter is the order
parameter exponent, defined as the long-time scaling of the density in
terms of the control parameter, $n(t \to \infty) \sim c^{\beta}$. By
restoring the diffusive term, the spatial exponent
$\nu_{\perp}=\frac{1}{2}$, defined analogously to $\nu_{\parallel}$,
may be identified.
 
In the following section we show that fluctuations alter these
predictions in low dimensions, by virtue of endowing space and time
scale exponents with small dimension-dependent corrections.  The
exponent $\beta$, however, remains unchanged. We argue that because
our model possesses detailed balance, which ensures that $\av{\phi(t
\to \infty)} \equiv \av{\phi} = c$, this fixes $\beta$ to unity. This
may also be inferred from the invariance of the unshifted action under
the transformation $\left(\phi,\ps \right) \to \left(c \ps, c^{-1}
\phi \right)$, and the consequent Ward identity~\cite{ZinnJustin}.

\subsection{Dimensional analysis}

\begin{figure}
\psfig{file=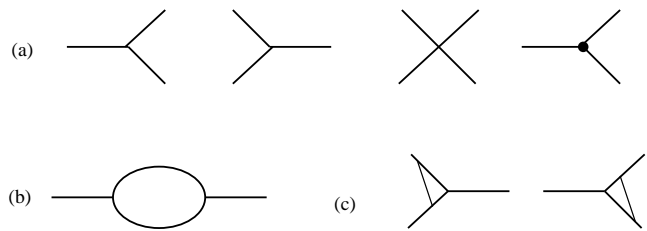,height=8.5cm,angle=270}
\caption{\label{fig:vertices} Vertices corresponding to terms in the
action (\ref{shift}). (a) From left to right, diagrams corresponding
to the vertices $-\lambda_0^{(1)}$, $v_0$, $-\lambda_0^{(2)}$, and
$\nu_0^{(1)} q_1^2$. The dot denotes a momentum dependence $q_1^2$ on
one of the incoming legs.  (b) The structure of the lowest-order
corrections to the propagator, showing how the effective couplings
emerge. The coupling shown here is $x$. Similar diagrams with dots on
right, left and both vertices correspond to propagator renormalization
controlled by the couplings $y$, $z$ and $u$, respectively. (c)
Diagrams renormalizing the coupling $x$. Because these diagrams are
symmetric with respect to incoming and outgoing momenta and
frequencies, the contribution from each is doubled. Note that (b) also
contributes to the renormalization of $x$, via the renormalization of
$D$.}
\end{figure}

We identify the upper critical dimension of the model via dimensional
analysis~\cite{Amit,Tauber}. We rescale space according to $x
\rightarrow x \rho^{1/2}$, in order to remove the temperature
dependence from the diffusion coefficient. Note that this rescaling is
not valid {\em at} $c=0$. The action (\ref{shift}) then reads as
before, with rescaled parameters $\lambda_0^{(1,2)} \equiv 2 d a^d
\rho^{-\frac{d}{2}}\gamma$, $r_0=v_0 \equiv 2d \rho$, $\sigma_0 \equiv
a^{2}$, and $\nu_0^{(1,2)} \equiv \gamma \rho^{-1-\frac{d}{2}}
a^{d+2}$. We show some of the diagrams corresponding to these
couplings in Fig.~\ref{fig:vertices}a.

To perform a scaling analysis, we identify the effective couplings
emerging from the action. These follow from the structure of the
diagram shown in Fig.~\ref{fig:vertices}b, and are
\begin{equation}
\label{couplings}
x_0 \equiv \frac{v_0 \lambda_0^{(1)}}{D_0^2}, y_0 \equiv
\frac{\sigma_0 \lambda_0^{(1)}}{D_0^2}, z_0 \equiv \frac{v_0
\nu_0^{(1)}}{D_0^2}, u_0 \equiv \frac{\sigma_0 \nu_0^{(1)}}{D_0^2}.
\end{equation}
The factors of $D_0$ come from the explicit evaluation of the
integrals associated with the diagrams. To (\ref{couplings}) we add
$g_0 \equiv \lambda_0^{(2)}/D_0$ and $h_0 \equiv \nu_0^{(2)}/D_0$,
which couple to four-point vertices: see
Fig.~\ref{fig:vertices}. Dimensional analysis reveals that the upper
critical dimension is 4, at which the most relevant coupling, $x_0$,
is marginal. Renormalization effects must therefore be taken into
account for $d<4$. Above $d=4$ the classical (fluctuation-free)
predictions apply. Other couplings become relevant below $d=2$, and we
shall therefore restrict our analysis to $2 \leq d \leq 4$. Dimension
$d=1$ is treated separately in Section~\ref{d1}.

\subsection{DP fixed point}

We employ the usual field-theoretic renormalization group
scheme~\cite{Amit,ZinnJustin}, using dimensional regularization in
$d=4-\epsilon$ dimensions to identify the unphysical ultra-violet (UV:
short time and distance) poles of the vertex functions
$\Gamma^{(\bar{N},N)}$ of the theory. The vertex functions
$\Gamma^{(\bar{N},N)}$ consist of all one-particle-irreducible
diagrams with $\bar{N}$ outgoing and $N$ incoming amputated
lines. Their UV poles result from exchanging a lattice model, which is
regularised at short distances, for a continuum field theory, which is
not. But by invoking universality, which says that the behaviour of a
system approaching criticality is governed by a small number of
relevant parameters, we recognise that the UV poles correspond to
irrelevant microscopic degrees of freedom. By removing these poles we
both render our theory finite, and, via scale-invariance and
dimensional analysis, infer its physically important infra-red (IR:
large time and distance) scaling~\cite{Tauber}. We shall work to
one-loop order, and use dimensional regularization and minimal
subtraction~\cite{Amit}.

We introduce the following renormalized counterparts of the fields and
couplings appearing in (\ref{shift}):
\begin{eqnarray}
\label{Z-factors-0}
\begin{array}{c}
\phi_R = Z_{\phi}^{-1} \phi,\quad \pb_R = \pb, \quad \lambda^{(1)} =
 Z_{\phi}^2 Z_{\lambda} \lambda_0^{(1)} \mu^{d-2}, \nonumber \\
 v=Z_{\phi} Z_{v} v_0 \mu^{-2},\quad r = Z_{\phi} Z_{r} (r_0-r_{0c})
 \mu^{-2}, \nonumber \\ D=Z_{\phi} Z_D D_0,
\end{array}
\end{eqnarray}
where $r_{0c}$ is the additive counterterm introduced to cure the
quadratic divergence of the vertex function $\Gamma^{(1,1)}$. We have
introduced an arbitrary momentum scale $\mu$ in order to render the
couplings dimensionless, and have chosen to allocate dimensions to the
fields according to $[\pb]=1, [\phi]=\mu^d$. The predictions of the
theory must be independent of this allocation.

We define the multiplicative renormalization factors $Z$ as
follows. From the propagator couplings, we fix mass, field and
diffusion constant renormalization via
\begin{eqnarray}
\label{rg-scheme}
\left. \partial_{i \omega} \goo_R(\omega,q) \right|_{NP} =1, \nonumber \\
 \left. \partial_{q^2} \goo_R(\omega,q)  \right|_{NP}=D, \nonumber\\  
-\goo_R(0,0)=r \mu^2,
\end{eqnarray}
while for the couplings comprising $x_0$ we impose the conditions
\begin{eqnarray}
\label{rg-scheme-vertices}
&&\lambda^{(1)}=\frac{1}{2} \left. \got_R(\omega,q)
\right|_{NP} \mu^{d-2}, \nonumber \\ &&v= -\frac{1}{2}\left.
\gto_R(\omega,q) \right|_{NP} \mu^{-2}.
\end{eqnarray}
The subscript $NP$ stands for `normalization point', and is the value
of the external momentum scale at which we evaluate the vertex
functions. It can be chosen for convenience, provided that it lies
outside the IR-singular region; we take $NP = (i\omega,q^2,r)=(2 D
\mu^2,0,0)$. Note that this choice corresponds to the system at
criticality, which for finite $T$ is an approximation. For non-zero
$T$ one must retain the mass term in the propagator. This leads to the
emergence of an effective coupling that flows logarithmically to zero,
signaling a crossover to a massive, classical fixed point. We will
discuss this case in Section~\ref{Tpos}.

We first assume that for $2<d\leq 4$ the couplings other than $x_0$
are irrelevant, and hence the action (\ref{shift}) reduces to that of
DP (we shall call $x$ the `DP coupling'). We shall find that those
couplings which are marginal in $d=2$ at the classical fixed point are
rendered irrelevant at the DP fixed point. Hence we expect to see DP
scaling for $2 \leq d \leq 4$. We find, to one-loop order, the
well-known $Z$-factors~\cite{Hinrichsen},
\begin{eqnarray}
\label{prop-Z}
Z_{\phi}=1+\frac{A_d}{4 \epsilon}x_0 \mu^{-\epsilon}, \quad
Z_{D}=1-\frac{A_d}{8 \epsilon}x_0 \mu^{-\epsilon}, \nonumber \\
Z_{r}=1-\frac{A_d}{2 \epsilon}x_0 \mu^{-\epsilon}, \quad
Z_{\lambda^{(1)}} = 1-\frac{A_d}{ \epsilon} x_0 \mu^{-\epsilon},
\nonumber \\ Z_{v} = 1-\frac{A_d}{ \epsilon} x_0 \mu^{-\epsilon},
\end{eqnarray}
where $A_d \equiv 4 (4 \pi)^{-d/2} \Gamma(3-d/2)$. Note that the cubic
vertices renormalize identically as a consequence of a Ward
identity. The renormalization factor associated with $x=\lambda^{(1)}
v/D^2$ is therefore $Z_x=Z_{\phi} Z_{\lambda^{(1)}} Z_v Z_D^{-2}= 1-(3
A_d/2 \epsilon) x_0 \mu^{-\epsilon}+ \cal{O}(\epsilon^2)$. Insofar as
one can ignore the propagator mass, the rescaled coupling $x
\rightarrow A_d x=A_d Z_x x_0 \mu^{-\epsilon}$ changes with the
observation scale $\mu$ according to
\begin{equation}
\label{beta-x}
\beta_x \equiv \mu \frac{\partial x}{\partial \mu} = x \left(-\epsilon
+\frac{3}{2}x\right).
\end{equation} 
If we parameterize the change in the observation scale by $\mu
\rightarrow \mu(\ell) =\mu \ell$, we can solve (\ref{beta-x}) for
$x(\ell)$:
\begin{equation}
\label{flow}
x(\ell) = \frac{x^{\star}}{ \left( \frac{ x^{\star}}{x(1)} -1 \right)
\ell^{\epsilon} +1}.
\end{equation}
Thus $x \rightarrow x^{\star} \equiv 2 \epsilon/3$ as $\ell
\rightarrow 0$, because $\epsilon >0$. Since $\ell \approx 1$ and
$\ell \ll 1$ correspond respectively to microscopic and macroscopic
length and time scales, $x^{\star}$ is an IR-stable fixed point. At
this fixed point the critical exponents of the theory are independent
of its microscopic parameters, and so are `universal'. We therefore
expect the model to display scaling behaviour independent of its
microscopic details for very low temperatures. This scaling behaviour
belongs to the universality class of directed percolation.

Having assumed the non-DP couplings in the action are irrelevant for
$2<d \leq4$, we shall now justify this assumption. These couplings are
indeed irrelevant at the classical fixed point, as one can verify from
(\ref{shift}) by dimensional analysis. We find that they remain
irrelevant at the DP fixed point. Further, those couplings $(y,z,g)$
which are marginal in $d=2$ at the classical fixed point are rendered
irrelevant at the DP fixed point. Hence we expect to see DP scaling in
$d=2$, also. Defining a renormalization scheme in a similar manner to
before,
\begin{eqnarray}
\label{rg-scheme-2}
\lambda^{(2)} = Z_{\phi}^2 Z_{\lambda^{(2)}} \lambda_0^{(2)}
 \mu^{d-2}, \quad \, \sigma = Z_{\phi} Z_{\sigma} \sigma_0 , \quad
 \nonumber \\ \nu^{(1)}= Z_{\phi}^2 \nu_0^{(1)} Z_{\nu^{(1)}} \mu^{d},
 \quad \, \nu^{(2)}= Z_{\phi}^2 Z_{\nu^{(2)}} \nu_0^{(2)} \mu^{d},
\end{eqnarray}
where 
\begin{eqnarray}
\label{rg-scheme-2b}
\lambda^{(2)}=\left. \frac{1}{4}  \gtt_R(q,\omega)
\right|_{NP} \mu^{d-2}, \nonumber \\ \sigma=\left. 
\partial_{q^2} \gto_R(q,\omega) \right|_{NP}, \nonumber \\
\nu^{(1)}=\left. \partial_{q^2} \got_R(q,\omega) \right|_{NP}
\mu^d, \nonumber \\ \nu^{(2)}=\left.  \partial_{q^2}
\gtt_R(q,\omega) \right|_{NP} \mu^d.
\end{eqnarray}
we find to one-loop order
\begin{eqnarray}
Z_{\nu^{(1)}} = Z_{\sigma} = 1-\frac{2 A_d}{ \epsilon} x_0
\mu^{-\epsilon}, \nonumber \\ Z_{\lambda^{(2)}} = 1-\frac{A_d}{
\epsilon} x_0 \mu^{-\epsilon}.
\end{eqnarray}
The corrections to the Gaussian scaling dimensions of these couplings
may then be calculated. The correction to the Gaussian eigenvalue of
$g \equiv \lambda^{(2)}/D$ is determined by $Z_g = Z_{\phi}
Z_{\lambda^{(2)}} Z_D^{-1} = 1- 5 x_0 \mu^{-\epsilon}/(8
\epsilon)$. So $\gamma_g^{\star} \equiv \left. \mu \partial_{\mu}
\ln(g/g_0) \right|_{x=x^{\star}}= d-2+5\epsilon/12$. Thus $g$ is less
relevant at the DP fixed point than at the Gaussian fixed point, and
for $d \geq 2$ may safely be ignored. So too may $h$. In a similar way
we find that $\gamma^{\star}_y=\gamma^{\star}_z=d-2+5 \epsilon/3$, and
$\gamma^{\star}_u=d+7\epsilon/3$, all of which are irrelevant for $d
\geq 2$ at the DP fixed point. Thus for $2 \leq d \leq 4$ the scaling
properties of the model near criticality are those of DP.

The critical exponents of the model then follow from standard
arguments~\cite{Amit,Hinrichsen}. They are, to $\cal{O}(\epsilon)$,
$\left( \nu_{\perp}^{DP},\nu_{\parallel}^{DP} \right) = \left(
\frac{1}{2}+\frac{\epsilon}{16},1+\frac{\epsilon}{12} \right)$. The
temporal exponent appropriate for comparing these predictions with
numerical simulations is $1+\nu_{\parallel}$. The additional factor of
unity arises because the microscopic timescale associated with the
system goes itself as $\sim c$. The non-trivial value of $\beta$,
$\beta^{DP}=1-\epsilon/6$, cannot be observed for a model such as ours
which eventually equilibrates, as discussed above.

\subsection{Critical temperature}

In general, systems in the DP universality class, such as directed
bond percolation in $1+1$ dimensions, exhibit a continuous phase
transition from an active to an absorbing state at some finite value
$p_c \neq 0$ of their control parameter $p$. Our model, for which $p =
c$, displays no such transition. One can justify this difference on
physical grounds, as follows. If we interpret the mobility $n_i$ as
the concentration of a chemical reactant $A$, then the KCM we study
for $d>2$ corresponds to a chemical reaction involving diffusion
$(A+\emptyset \leftrightarrow \emptyset+A)$, branching $(A+\emptyset
\rightarrow A+A)$ and coagulation $(A+A\rightarrow
A+\emptyset)$. Recall that the diffusive process arises from the
mechanism of mobility creation facilitated by a nearest-neighbour
site, and is made manifest only following a shift of the response
field. DP corresponds to these three processes {\em plus} self
destruction, $A+\emptyset \stackrel{\sigma_0}{\longrightarrow}
\emptyset+\emptyset$. It is self destruction that permits other
systems in the DP universality class to undergo a phase transition at
a finite value of the control parameter. Self destruction gives rise
to a second-quantized operator
\begin{equation}
H_{sd}=-\sigma_0 (a_i-\ad_i a_i),
\end{equation}
which, following a shift of the response field, results in a term in
the action of the form $\sigma_0 \pb \phi$. Thus the mean-field
critical point becomes $p_c = \sigma_0$. Near criticality, $p_c$ is
increased above its mean-field value by fluctuations. This occurs
because the DP noise-noise correlator is positive, and so coagulation
is enhanced by the branching process: each particle finds itself with
more neighbours with which it may coagulate than one would expect from
a mean-field approximation. This enhanced coagulation enters the term
which renormalizes the mass, effectively enhancing self-destruction
relative to branching, and shifting the critical percolation threshold
upwards.

Now self destruction is excluded by any dynamical rule preventing
mobility destruction unless facilitated by a nearest
neighbour. Moreover, no such process can be generated under
renormalization from only branching and coagulation processes whose
respective rates are fixed by detailed balance. Hence we expect
one-spin facilitated models in general to have a critical point at
zero temperature.

This argument may be made explicit for the model we study. By imposing
the condition for criticality, $0=\goo(\omega=0,q=0,r_0=r_{0c})$, we
find that, to one-loop order
\begin{equation}
\label{mass-shift-cutoff}
r_{0c} = \frac{\left(\lambda_0 v_0/D_0 \right) \tilde{N}_d
\Lambda^{d-2}}{1-\left(\lambda_0 v_0/D_0^2 \right) N_d \Lambda^{d-4}}.
\end{equation}
Here, for convenience, we have imposed an explicit wavevector cutoff
$\Lambda$; the additive correction to the mass is formally equal to
zero in dimensional regularization, and yet the physical shift of the
critical temperature must be independent of the regularization scheme
used~\cite{Tauber}. We have introduced $N_d (2 \pi)^{-d} (d-4)^{-1}
\cal{S}_d$, where $\cal{S}_d \equiv 2 \pi^{d/2}/\Gamma(d/2)$ is the
surface area of a $d$-dimensional hypersphere, and $\tilde{N}_d \equiv
(d-4) N_d /(d-2)$. We also use the unscaled variables of the
action~(\ref{shift}), in which $D_0 \propto c$.

From (\ref{mass-shift-cutoff}) we see that the critical bare mass
changes sign as $T \rightarrow 0$ from above. Ostensibly the critical
temperature is then negative; physically, of course, it is zero. This
is a consequence of the vanishing of fluctuations in the limit of zero
temperature, which may be inferred from the vanishing in that limit of
the branching vertex in the action. The diffusion term arises from
nearest-neighbour-facilitated branching, and so must also vanish in
this limit. Thus there is no fluctuation-induced shift of the critical
temperature which remains $T_c=0$.

This is as we expect, if the field theory is a faithful representation
of the original master equation. The master equation satisfies
detailed balance at all temperatures, which means that it cannot admit
an absorbing state: an absorbing state breaks detailed balance because
it is a state that may be entered, but not left. Nonetheless, it is
necessary to verify, as in Eq.~(\ref{mass-shift-cutoff}), that there
exists no finite-temperature absorbing state under coarse-graining of
the master equation. The FA model, upon which the field theory is
based, is known to have a critical point at zero
temperature~\cite{Ritort-Sollich}.

\subsection{Crossover to classical behaviour}
\label{Tpos}

For any $T>0$ the mass parameter $r_0 \propto c$ will be
non-zero. Under renormalization, as discussed above, it will
eventually become large, rendering our approximation of criticality
incorrect. The system will thus for very large time and length scales
exhibit classical scaling properties, with the associated simple
exponents.

We can quantify the emergence of the classical theory by retaining the
mass term in the propagator~\cite{Cardy-et-al}. If we write $s \equiv
r/D$, we find that
\begin{eqnarray}
\label{ex}
 \frac{d  \ln x(\ell)}{d \ln \ell} &=   -\epsilon +\frac{3}{2} g(\ell),\\
\label{ess}
 \frac{d  \ln s(\ell)}{d \ln \ell} &=   -2 +\frac{3}{8} g(\ell) ,
\end{eqnarray}
where $g(\ell) \equiv x(\ell) (1+s(\ell))^{d/2-2}$ is an effective
coupling. For small $s$, $x$ would in the IR limit approach to the
directed percolation fixed point. But $s$ does not remain small,
flowing as $s(\ell) \sim \ell^{-2 + \mathcal{O}(\epsilon)}$. If we
introduce the scaled mass $\sigma \equiv \left(1+s(1)^{-1}
\ell^{2+\cal{O}(\epsilon)}\right)^{-1}$, we find that in the large
mass limit $\sigma \to 1$ we obtain a logarithmically diminishing
coupling,
\begin{equation}
g(\ell) \sim  g(1) \left\{ 1+ \frac{3}{16}(4+d) g(1) \ln \ell +
\cdots \right\}^{-1}.
\end{equation}
The vanishing of the effective coupling signals the re-emergence of a
classical theory: because $g$ couples to diagrams renormalizing the
propagator, its logarithmic vanishing results in a logarithmic
crossover to classical exponents.

Thus we should see DP scaling provided that temperatures are small
enough and time and length scales are not too large. The crossover
temperature will be system dependent, because the prefactors of the
flowing couplings are non-universal. For larger temperature or large
enough length and time scales we expect to see a logarithmic crossover
to a classical theory. This is characterised, in the nonequilibrium
regime, by exponential decay to the steady state, and in general by
classical scaling behaviour.

\section{Dimension $d=1$ and CDP}
\label{d1}

For $d=1$ DP scaling no longer holds. This is signaled in the field
theory by the relevance of some of the non-DP couplings between $d=2$
and $d=1$, and the resulting profusion of uncontrollable
singularities~\cite{Amit}.  In this section we argue that in $d=1$
systems with single-spin isotropic facilitation, such as the FA model,
belong instead to the universality class of compact directed
percolation (CDP)~\cite{Hinrichsen}.

\begin{figure}
\begin{center}
\psfig{file=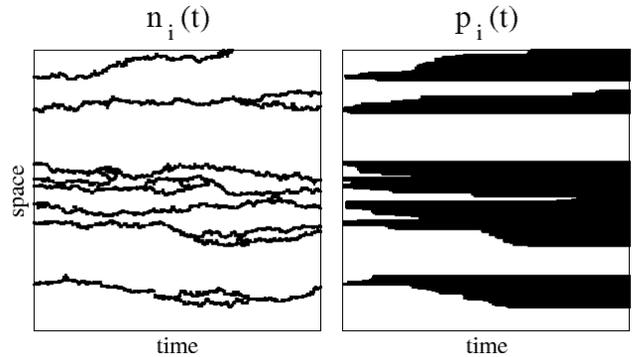,width=8.5cm}
\caption{\label{fig:CDP}FA model in $d=1$. The left panel shows an
equilibrium trajectory for the mobility field $n_i(t)$ at $T=0.3$
(window size is $L=250$ by $\Delta t=5000$). The right panel shows the
corresponding trajectory of the persistence field $P_i(t)$; black
sites denote those which have been or are mobile, and so satisfy
$P_i(t)=0$. The clusters generated by the dynamics of the $P_i(t)$ are
compact, and their scaling properties are that of CDP.}
\end{center}
\end{figure}

Consider the FA model in $d=1$.  The elementary order parameter of
this model is the mobility field $n_i(t)$.  Figure \ref{fig:CDP} (left
panel) shows a portion of an equilibrium trajectory at $T=0.3$. The
connection between the FA model and CDP is made apparent by
considering instead the corresponding persistence field $P_i(t)$,
i.e. the field which takes value $P_i(t)=0$ if site $i$ has flipped by
time $t$, and $P_i(t)=1$ otherwise.  The corresponding trajectory of
$P_i(t)$ in our example is shown in Fig.~\ref{fig:CDP} (right panel).

Clearly, while the dynamics of $n_i(t)$ is reversible, that of
$P_i(t)$ is not. A related observation is that the clusters generated
by the evolution of $P_i(t)$ are compact, as seen in Fig.\
\ref{fig:CDP}. The control parameter is again $c$, with $c=0$
corresponding to the transition between an active phase in which
$P_i(t)$ eventually becomes unity throughout the whole system, and an
inactive phase, in which it does not. As before, the exponents
$\nu_{\parallel}$ and $\nu_{\perp}$ determine the scaling of times,
$(D \tau) \sim c^{-\nu_{\parallel}}$ (with $D \approx c$), and
lengths, $\xi \sim c^{-\nu_{\perp}}$. Two further exponents determine
the asymptotic values of $\langle P_i(t) \rangle$. To extract these
exponents it is convenient to define the {\em transience} function
$T_i(t) \equiv 1-P_i(t)$: starting from an initial finite seed,
$\langle T_i(t \to \infty) \rangle \sim c^{\beta}$, with the dynamics
running in the forward time direction; starting from a completely full
lattice, $\langle T_i(t \to - \infty) \rangle \sim c^{\beta'}$, with
the dynamics running backwards in time.  Note that $\beta \neq \beta'$
due to the irreversibility of $T_i(t)$ (or $P_i(t)$).

The domains of $T_i(t)$ spread only through diffusion of mobility
excitations and interactions play no role.  In this sense, the scaling
behaviour of $T_i(t)$ should be that of freely-diffusing domain walls,
and coalescing domains. Examples of systems which behave similarly are
the zero temperature Ising chain under Glauber
dynamics~\cite{Hinrichsen}, or the reaction-diffusion system $A+A \to
\emptyset$~\cite{Peliti,Hinrichsen}. These indeed belong to the CDP
universality class.

CDP has the following exponents \cite{Hinrichsen}:
\begin{equation} 
\nu_{\parallel}^{\rm CDP} = 2, \;\;\; \nu_{\perp}^{\rm CDP} = 1,
\;\;\; \beta^{\rm CDP}=0, \;\;\; \beta'^{\rm CDP}=1 \; .
\label{CDP}
\end{equation}
These are precisely the values of the exponents of the FA model in
$d=1$. The time and length exponents are $\nu_{\parallel}^{\rm FA} =
2$ and $\nu_{\perp}^{\rm FA} = 1$, giving the dynamic exponent $z^{\rm
FA} \equiv \nu_{\parallel}^{\rm FA}/\nu_{\perp}^{\rm
FA}=2$~\cite{Garrahan-Chandler,Ritort-Sollich}. Each site of a lattice
which initially contains at least one excitation will eventually flip,
and thus $T_i(t \to \infty) = 1$ for all $i$, independently of $c$. We
therefore have $\beta^{\rm FA} = 0$, which is a consequence of
ergodicity in the active phase. Conversely, if one takes a final state
with all $T_i = 1$, and runs time backwards, the state at $t \to
-\infty$ will have a density of excitations, and therefore of $T_i$,
equal to $c$. This is a consequence of detailed balance. Hence
$\beta'^{\rm FA} = 1= \beta'^{\rm CDP}$.

We propose a field-theoretic justification for this behaviour as
follows. The Langevin equation of motion for $\phi$ is given by
(\ref{lang-eqn}) and (\ref{nn-corr}). At and above $d=2$ the term in
$\lambda^{(2)}$ is irrelevant at the DP fixed point and may be
dropped, leving us with the DP Langevin equation~\cite{Hinrichsen}. In
$d=1$, however, at the DP fixed point (assuming it exists), we have
from our previous results the anomalous dimensions of the couplings
appearing in the noise correlator:
\begin{equation}
\gamma_{\lambda^{(2)}}^{\star}=d-2+\epsilon/3, \quad
\gamma_v^{\star}=0.
\end{equation}
We have calculated these dimensions using the prescription
$[\pb]=[\phi]=\mu^{d/2}$, appropriate when the cubic vertices are
considered independently. We see that $\lambda^{(2)}$ and $v$ are both
marginal in $d=1$. This is, we stress, a crude approximation, because
the calculation of the anomalous dimensions assumes the irrelevance of
$\lambda^{(2)}$. But it does suggest that here this assumption is
inconsistent. Assuming that we can trust these exponents, there should
then exist a fixed point controlled by $\lambda^{(2)}$, at which $v$
is irrelevant. Assuming this is so, and assuming further that the
system can access this fixed point, this would leave the only vertices
in the effective theory $\pb \phi^2$ and $\left(\pb \phi \right)^2$,
which allow no propagator renormalization. Hence $z=2$ exactly. For
this theory the beta function is calculable to all orders, since
perturbation theory in $\lambda^{(2)}_0$ gives us a geometric
series~\cite{Peliti}. We then have a new fixed point, at which there
exists a renormalized value of $\lambda^{(1)}$ corresponding to an
infinite value of its bare counterpart, $\lambda_0^{(1)}$. Thus
$\lambda_0^{(1)} \phi^2 \gg r_0 \phi$, giving the effective theory
\begin{equation}
\partial_t \phi(t)= D_0 \nabla^2 \phi -\lambda_0^{(1)} \phi^2 + i
\sqrt{\lambda_0^{(2)}} \eta(x,t).
\end{equation}
This is the Langevin equation for the CDP universality
class~\cite{Peliti}. We stress that this argument is conjecture only.
A more rigorous analysis of the field theory would be required in
order to justify this claim.

\section{Simulations of the $d=3$ FA model}

The one-spin facilitated FA
model~\cite{Fredrickson-Andersen,Ritort-Sollich} is the lattice model
upon which the field theory of the previous sections is based. In this
section we report the results of our large-scale numerical simulations
of the equilibrium dynamics of the FA model in dimension $d=3$, and
compare these results to the predictions of the field theory. While
the one dimensional FA model has been extensively studied by numerical
simulations~\cite{Ritort-Sollich}, we are not aware of any detailed
numerical study for $d>1$.

We consider the FA model on a cubic lattice with periodic boundary
conditions. The model is defined by the Hamiltonian
(\ref{hamiltonian}), and the isotropic dynamical rule
\begin{equation}
n_i=0
\begin{array}{c}
\xrightarrow{~ ~ ~{\cal C}_i ~ c ~ ~ } \\ \xleftarrow[{\cal C}_i ~
(1-c)]{} \\
\end{array}
n_i=1 .
\label{rates}
\end{equation}
The kinetic contraint is ${\cal C}_i = 1-\prod_{\langle j,i \rangle}
(1-n_j)$, where $\langle j,i \rangle$ denotes nearest-neighbour pairs.
We perform Monte-Carlo simulations of this model for several
temperatures in the range $T \in [0.09, 5.0]$. We use the continuous
time algorithm~\cite{Newman-Barkema}, which is well-suited to this
problem. The dynamical slow-down in this model is accompanied by the
growth of a dynamic correlation length, and hence we must account for
possible finite size effects. For instance, at $T=0.09$ it was
necessary (and perhaps even then not sufficient: see below) to use
system sizes as large as $N=160^3$.

\subsection{Global dynamics}

\begin{figure}
\psfig{file=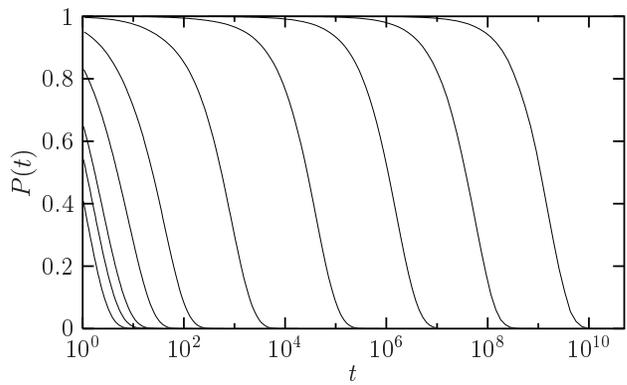,width=8.5cm}
\caption{\label{pers} Persistence function $P(t)$ for the $d=3$ FA
  model. From left to right: $T=5.0$, 1.5, 1.0, 0.6, 0.4, 0.25, 0.17,
  0.13, 0.106 and 0.009.}
\end{figure}

We first consider the spatially averaged dynamics. This may be probed
via the mean persistence function,
\begin{equation}
P(t) = \left\langle \frac{1}{N} \sum_i
P_i(t) \right\rangle,
\end{equation}
where $P_i(t)$ is the single-site persistence function at time $t$,
which takes value $1$ if site $i$ has not flipped up to time $t$, and
value $0$ otherwise. Fig.~\ref{pers} shows, as expected, that the
dynamics slows down markedly when temperature is decreased below $T_o
\approx 1.0$, which marks the onset of slow dynamics in this
model~\cite{Berthier-Garrahan,Brumer-Reichman}.

\begin{figure}
\psfig{file=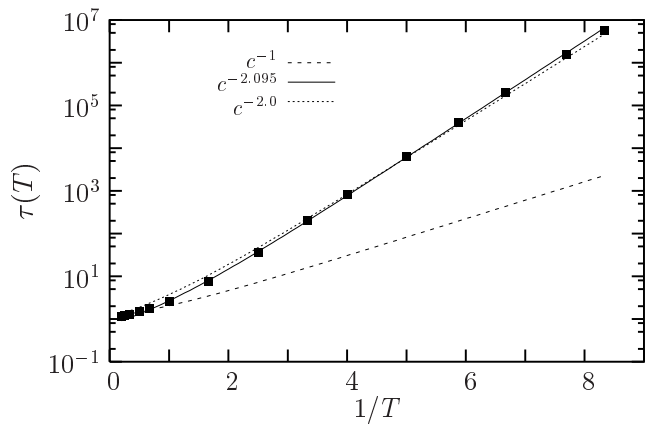,width=8.5cm}
\caption{\label{tauT} Arrhenius plot of the mean relaxation time in
the $d=3$ FA model. The DP exponent fits the data, while the classical
exponent does not.}
\end{figure}

We extract the mean relaxation time, $\tau(T)$, via the usual relation
$P(\tau) = e^{-1}$. The temperature dependence of $\tau$ is shown in
Fig.~\ref{tauT}, where various fits are also included. The high
temperature behaviour is well described by a naive mean-field
approximation~\cite{Berthier-Garrahan},
\begin{equation} 
\tau_{MF} \sim c^{-1}.
\end{equation}
This behaviour breaks down below $T_o$, where fluctuation-dominated
dynamics becomes important. From our field theoretic arguments we
expect that in the non-trivial scaling regime
\begin{equation}
\tau \sim c^{-\Delta}, \quad \Delta = 1+\nu_{\parallel} 
\approx 2.1, 
\end{equation}
where the numerical value is the DP estimate 
in three dimensions~\cite{Hinrichsen}. 
Fitting our data with the form $\tau \sim
c^{-\Delta}$ we find 
\begin{equation}
\Delta = 2.095 \pm 0.01 ,
\end{equation}
as shown in Fig.~\ref{tauT}. We include for comparison a fit using the
Gaussian value of the exponent, $\Delta=2$, which is inconsistent with
our data.

\begin{figure}
\psfig{file=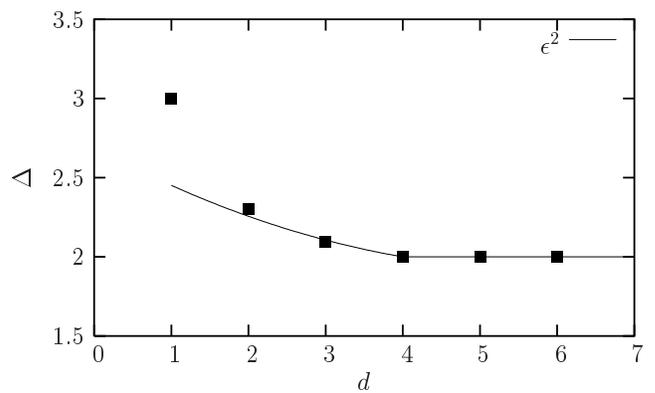,width=8.5cm}
\caption{\label{dpexp} Dimensionality dependence of the time exponent
$\Delta=1+\nu_{\parallel}$ of the FA model, from numerical
simulations in dimensions $d=1$ to 
$d=6$ (filled squares). 
The full line is the $\epsilon^2$ expansion prediction.
The values of the exponents agree within errorbars with those for DP
for all $d > 1$, and and CDP for $d=1$. 
The upper critical dimension of the
FA model is $d_c^{FA} = d_c^{DP} =4$.}
\end{figure}

We show in Fig.~\ref{dpexp} the results for similar simulations of the
FA model in dimensions $d$ from $1$ to $6$, together with the relevant
DP exponent to $\cal{O}(\epsilon^2)$, as tabulated in
Ref.~\cite{Hinrichsen}. We note that these results are consistent also
with numerical simulations of systems in the DP universality
class~\cite{Hinrichsen}. Our numerics also show that one-spin
facilitated FA models display, in all dimensions, Arrhenius
behaviour. They are thus coarse-grained models for strong
glass-formers, as expected~\cite{Garrahan-Chandler-2}.

In summary, Fig.~\ref{dpexp} strongly 
supports the RG prediction that the FA
model exhibits non-classical scaling in low dimensions, consistent
with DP behaviour for $d \geq 2$, and CDP behaviour in $d=1$.

\subsection{Distribution of relaxation times}

\begin{figure}
\psfig{file=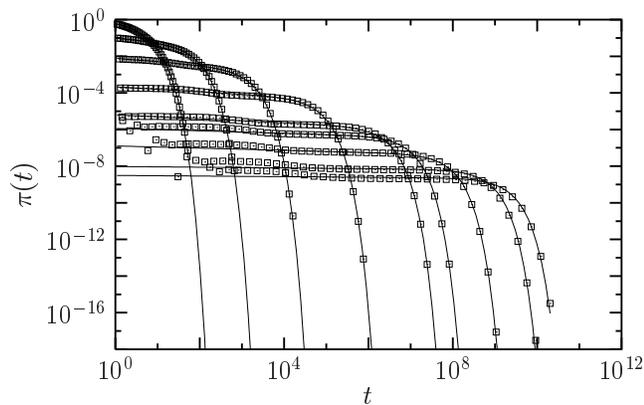,width=8.5cm}
\caption{\label{distrib} Distribution of relaxation times (symbols)
and fits to Eq.~(\ref{dis-fit}) (full lines) 
for temperatures $T=1.0$, 0.4, 0.25, 0.17, 0.13,
0.12, 0.106, and 0.09 (from left to right).}
\end{figure}

The mean relaxation time $\tau(T)$ captures only in part the
relaxation behaviour of the model. We consider in this subsection the
distribution of relaxation times, $\pi(t)$, related to the mean
persistence function via~\cite{Berthier-Garrahan}
\begin{equation}
P(t) = \int_t^\infty dt' \pi(t').
\end{equation}
These distributions are shown in Fig.~\ref{distrib}.

A careful study of the functions $P(t)$ and $\pi(t)$ reveals the
following structure. At very large times, the persistence decays to 0
in a purely exponential manner, $P(t \gg \tau) \sim \exp (-t/\tau)$.
This is not the case in $d=1$, where asymptotically the decay is
described by a stretched exponential with stretching exponent $\beta =
1/2$. That stretched exponential behaviour is not seen in $d=3$ is
consistent with the fact that strong glass-formers display an
almost-exponential relaxation pattern~\cite{Angell}.

\begin{figure}
\psfig{file=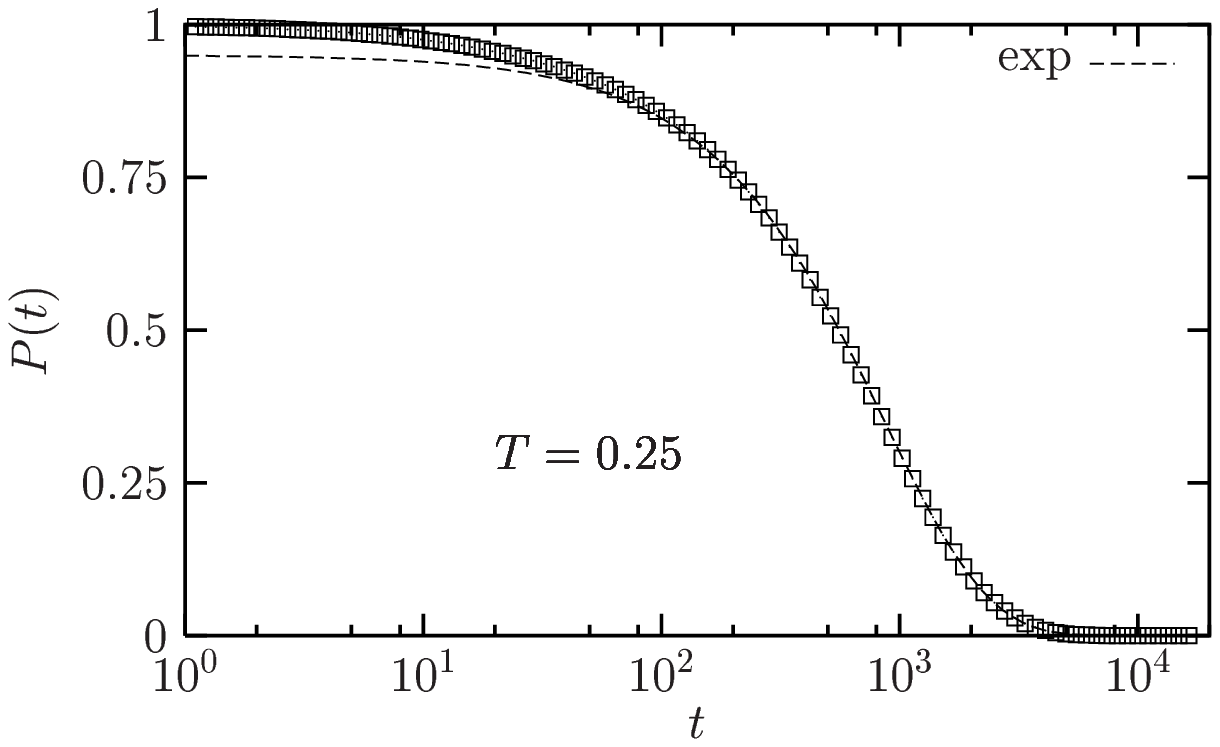,width=8.5cm}
\psfig{file=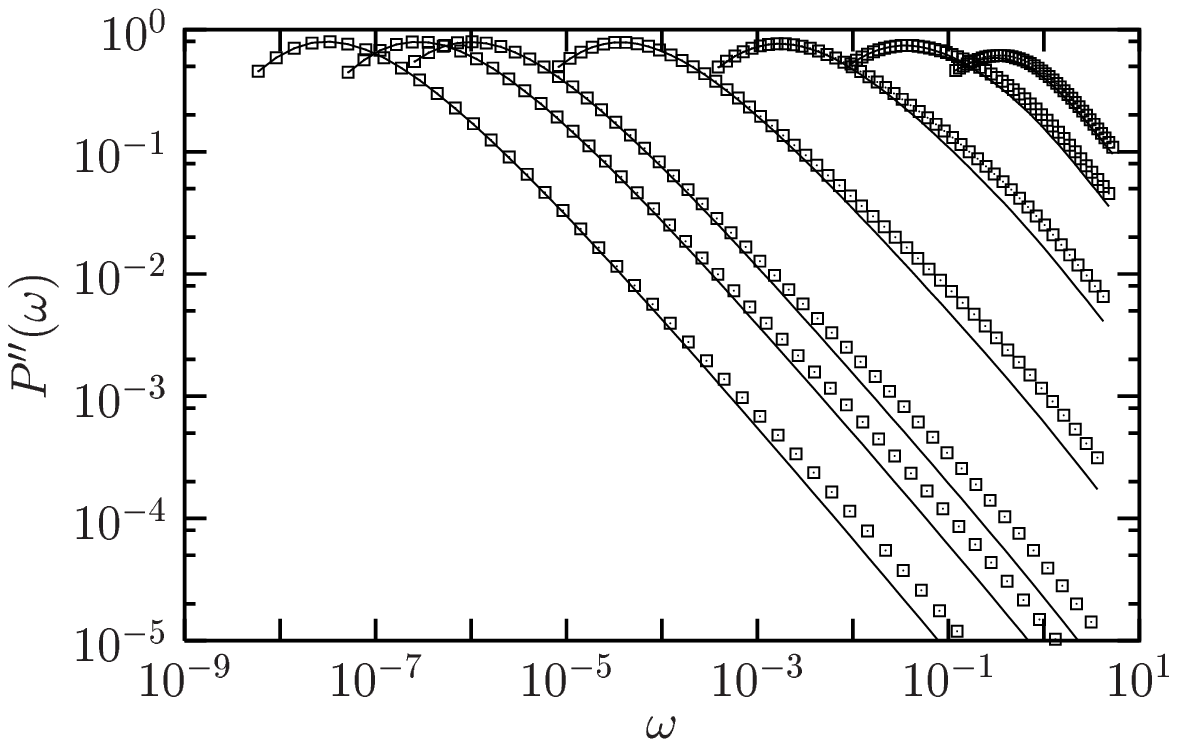,width=8.5cm}
\caption{\label{wing} Top: The fit of the persistence function with a
simple exponential reveals an additional short-time process.  Bottom:
This is also true in Fourier space, where the additional process looks
like a `high-frequency wing'.}
\end{figure}

Using $\tau(T)$ as a unique fitting parameter does not allow a
satisfactory description of the whole decay of the persistence
function: see Fig.~\ref{wing}. This figure shows that there exists an
`additional short-time process', in the language of 
glass transition dynamical studies.

Indeed, we find that fitting our data with the expression
\begin{equation}
\pi(t) \sim t^{-a} \exp \left( - \frac{t}{\tau} \right),
\label{dis-fit}
\end{equation} 
where $a$ and $\tau$ are free parameters, describes the distributions
reasonably well over several decades: see Fig.~\ref{distrib}.

Often, data in the supercooled liquid literature are presented in the
frequency domain, because many decades can be accessed via
e.g. dielectric spectroscopy~\cite{nagel}. 
Following this convention, we present
frequency data obtained from the distribution of time scales via
\begin{equation}
P(\omega) = \int_{-\infty}^\infty 
 \pi(\log(\tau)) \frac{1}{1+i \omega \tau} \, d \log \tau.
\end{equation}
Interestingly, the short-time power law behaviour $\pi(t) \sim t^{-a}$
observed in the distributions of time scales is also apparent in the
frequency space as an `additional process' on the high-frequency flank
of the $\alpha$ relaxation, $P''(\omega) \sim \omega^{-1+a}$: see
Fig.~\ref{wing}.  In this figure, the full lines correspond to fits of
the main peak with a simple exponential, as discussed above.

This feature is reminiscent of the `high-frequency wing' discussed at
length in the dielectric spectroscopy literature~\cite{nagel}.  The
wing is usually oberved in fragile glass-formers; unfortunately, no
dielectric data is available for strong glass-formers~\cite{leheny}.
Other techniques, such as Photon Correlation Spectroscopy, hint at the
presence of an additional process in strong glass-formers similar to
that observed in Fig.~\ref{wing}~\cite{strong}. More experimental
studies of the dynamics of strong glass-formers would 
be needed to confirm and quantify this similarity.

\subsection{Dynamic heterogeneity}

\begin{figure}
\psfig{file=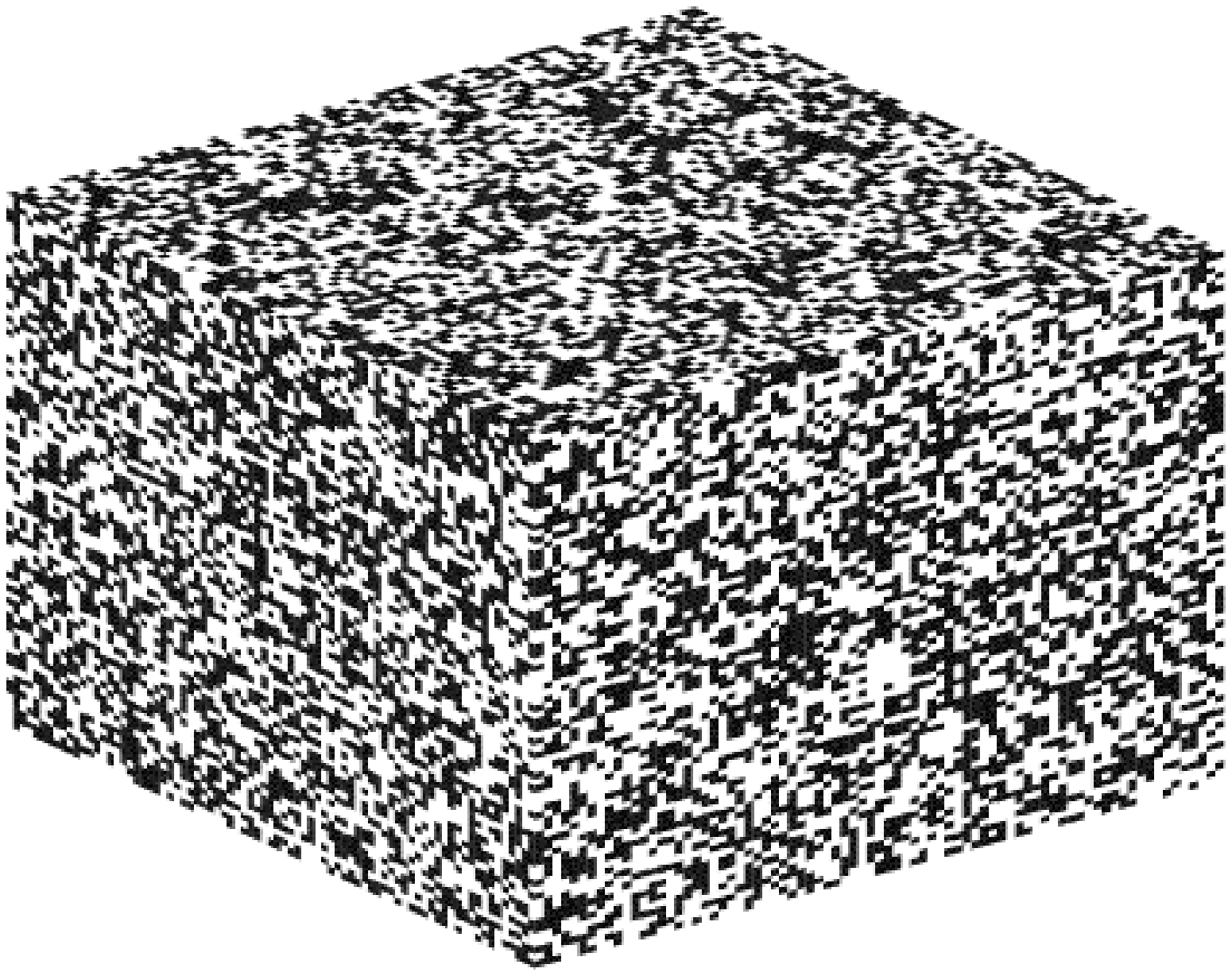,width=6.cm,height=6.cm}
\psfig{file=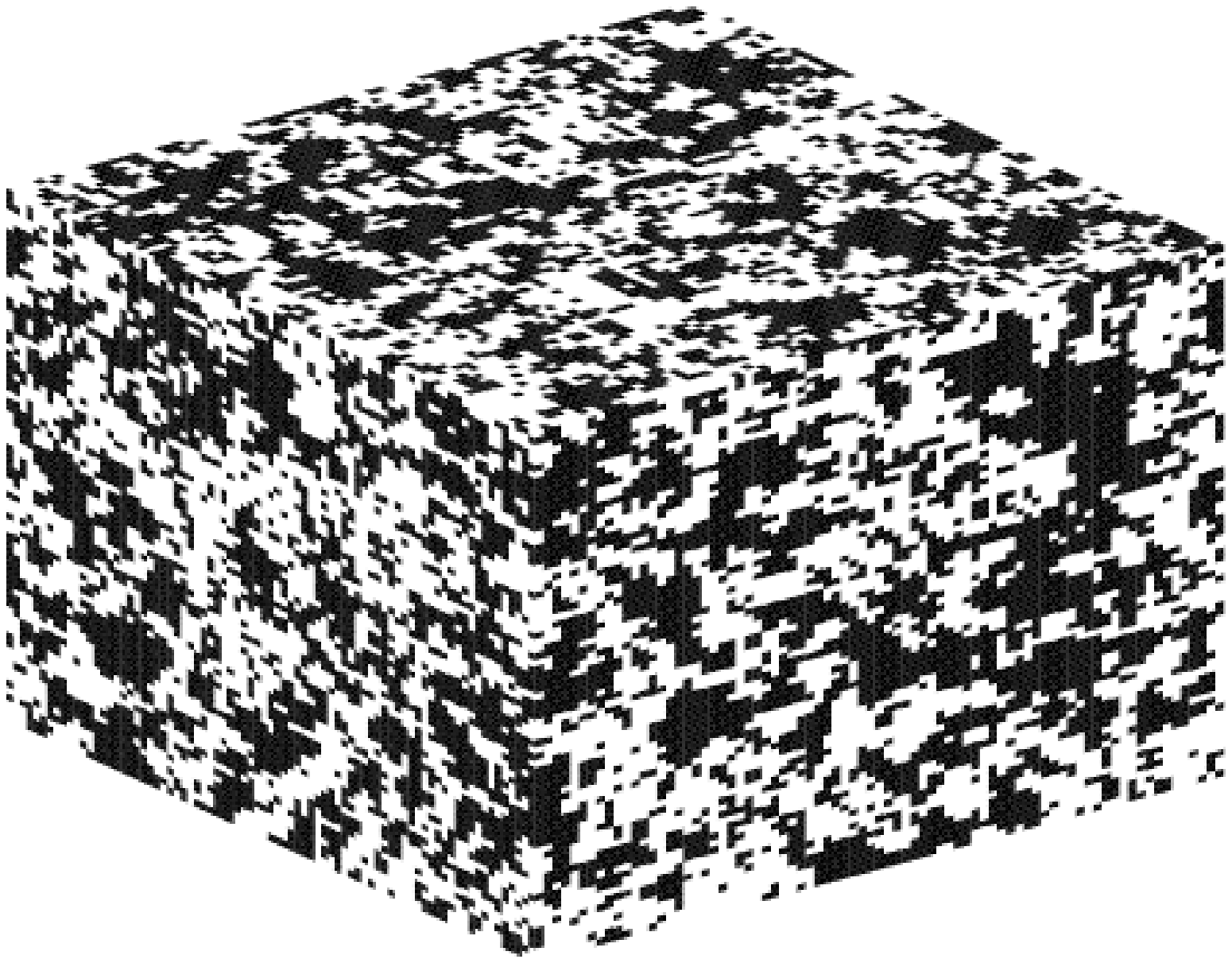,width=6.cm,height=6.cm}
\psfig{file=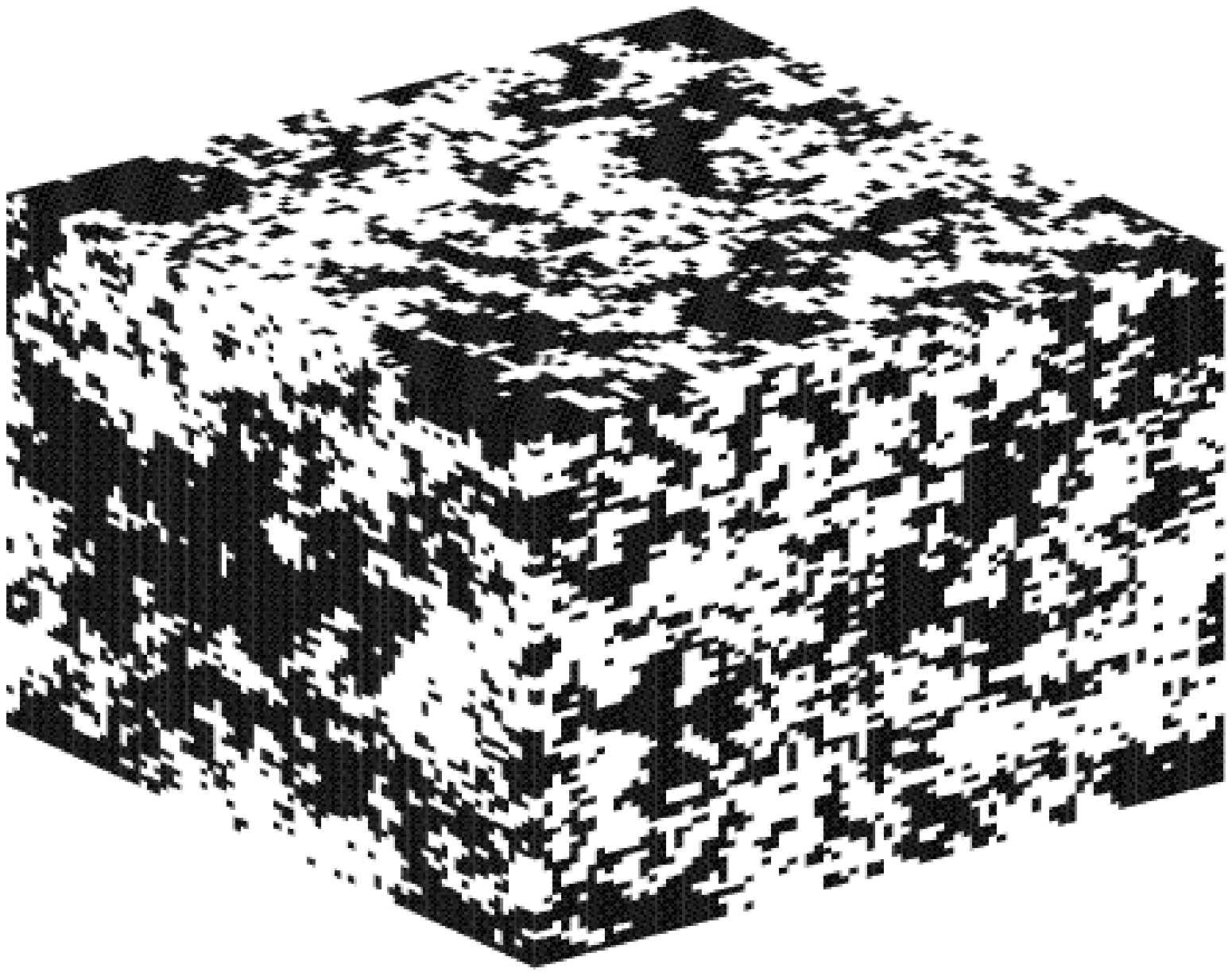,width=6.cm,height=6.cm}
\caption{\label{bubbles} Spatial distribution of the local 
persistence at time $t^\star$ such that $P(t^\star)=1/2$ (i.e., 50\%
of sites, shown in black, have flipped by time $t^\star$). 
From top to bottom, $T=1.0$, 0.2 and 0.12.
The appearance of dynamic critical fluctuations
when $T \to 0$ is evident.}
\end{figure}

The growth of timescales in the FA model, $\tau \sim c^{-\Delta}$, is
accompanied by growing spatial correlations, $\xi \sim
c^{-\nu_{\perp}}$, as the system approaches its critical point at
$T=0$. These correlations are purely dynamical in origin, and give
rise to dynamic heterogeneity~\cite{Garrahan-Chandler,Berthier-et-al}.
Figure \ref{bubbles} illustrates this phenomenon in the FA model. We
quantify the local dynamics via the persistence function $P_i(t)$.
For a given temperature we run the dynamics for a time $t^\star$, such
that $P(t^\star) = 1/2$, meaning that half of the sites have flipped
at least once. We colour white persistent (immobile) spins, for which
$P_i(t^\star)=1$, and black transient (currently or previously mobile)
spins, for which $P_i(t^\star) = 0$. Figure~\ref{bubbles} shows the
persistence function for the $d=3$ FA model at different
temperatures. Clearly, the dynamics is heterogeneous, and the spatial
correlations of the local dynamics grow as $T$ is decreased. The
`critical' nature of dynamic clusters is apparent: the pictures are
reminiscent of the spatial fluctuations of an order parameter close to
a continuous phase transition, such as the magnetization of an Ising
model near criticality. In our case, the order parameter is a dynamic
object, the persistence function, and the critical fluctuations are
purely dynamical in origin~\cite{fss}.

\begin{figure}
\psfig{file=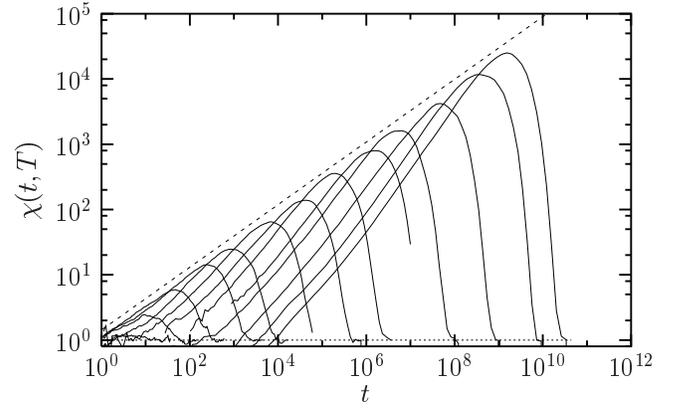,width=8.5cm}
\caption{\label{chi} Time dependence of the dynamic susceptibility
(\ref{chichi}) at different temperatures. From left to right $T=1.0$,
0.6, 0.4, 0.3, 0.25, 0.2, 0.17, 0.15 and 0.13. The horizontal dotted
line denotes the infinite time value, $\chi(t \to \infty) =1$; the
diagonal dotted line denotes the power-law fit $\chi(\tau,T) \sim
\tau^{0.46}$.}
\end{figure}

We now quantify these observations.  We can measure spatial
correlations of the local dynamics via a spatial correlator of the
persistence function,
\begin{equation}
\label{c}
C(r,t,T) = \frac{1}{N f(t)}\sum_i \Big[ \langle P_i(t) P_{i+r}(t)
\rangle - P^2(t) \Big],
\end{equation}
where the function $f(t) = P(t)-P^2(t)$ in the denominator ensures the
normalization $C(r=0,t,T)=1$.  Alternatively, one can take the Fourier
transform of (\ref{c}), giving the corresponding structure factor of
the dynamic heterogeneity,
\begin{equation}
\nonumber 
S(q,t,T) = \frac{1}{N f(t)} \sum_{k,l} \Big[ \langle P_k(t)
P_l(t) \rangle - P^2(t) \Big] e^{i q ( k - l )}.
\end{equation}
Finally, the zero wavevector limit of $S(q,t)$ 
defines a dynamic susceptibility,
$\chi(t,T) = S(q=0,t,T)$, which can be rewritten as the normalized
variance of the (unaveraged) persistence function, $p(t) \equiv 
N^{-1} \sum_i P_i(t)$:
\begin{equation}
\label{chichi}
\chi(t,T) = \frac{N}{f(t)} \big[ \langle p^2(t) \rangle - \langle p(t)
\rangle^2 \big].
\end{equation}

Figure~\ref{chi} shows the time dependence of the susceptibility
(\ref{chichi}) for various temperatures. The behaviour of $\chi$ is
similar to that observed in atomistic simulations of supercooled
liquids in general~\cite{DHnum}, and strong liquids in
particular~\cite{Vogel-Glotzer}. The susceptibility develops at low
temperature a peak whose amplitude increases, and whose position shifts to
larger times as $T$ decreases.  As expected, the location of the peak
scales with the relaxation time $\tau(T)$, indicating that dynamical
trajectories are maximally heterogeneous when $t \approx \tau(T)$.
\begin{figure}
\psfig{file=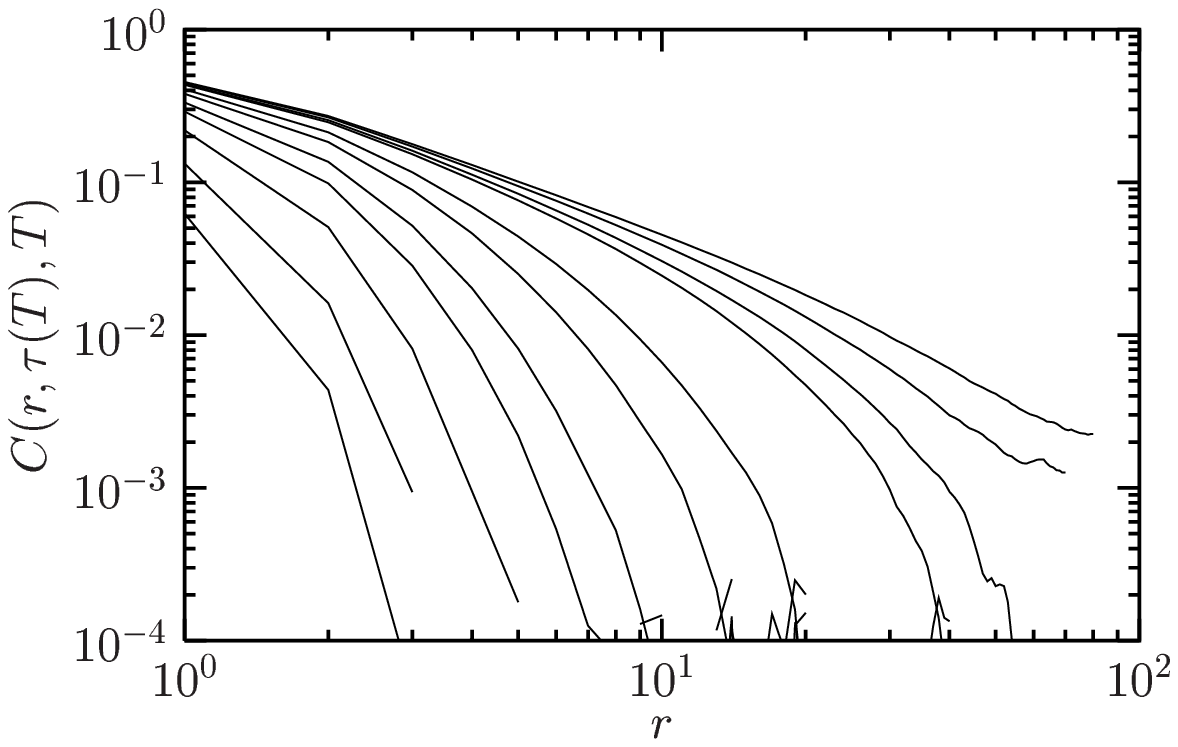,width=8.5cm}
\psfig{file=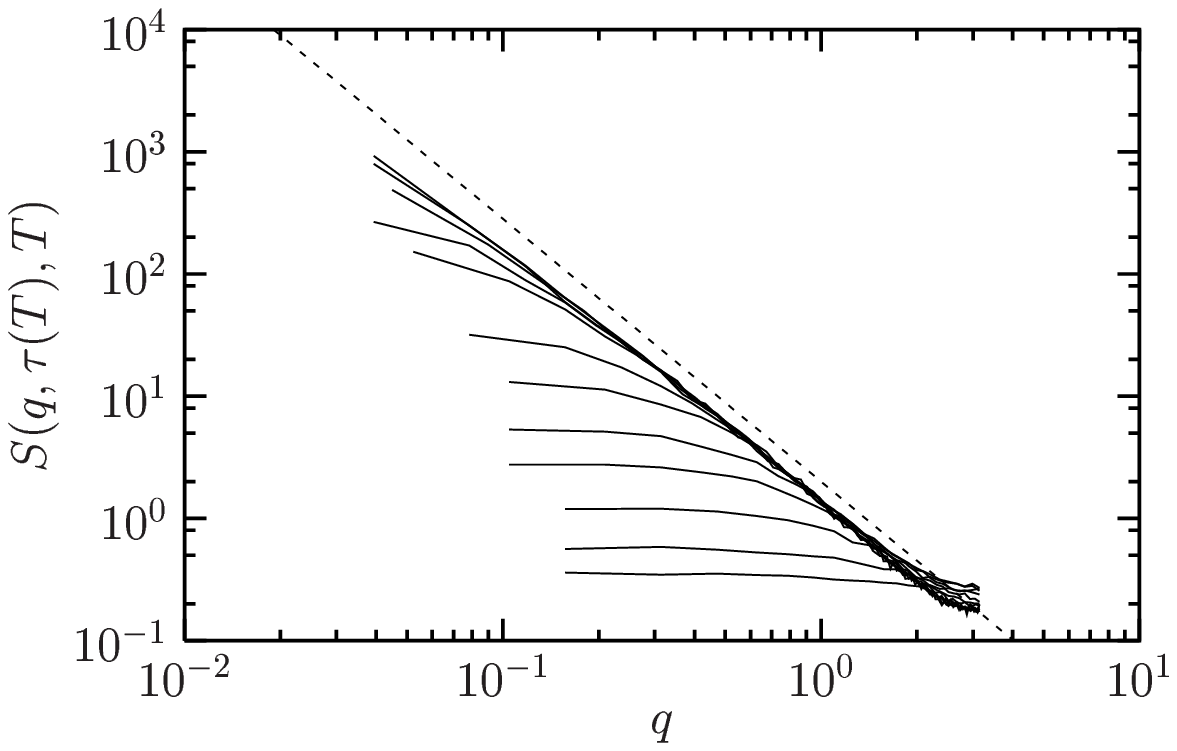,width=8.5cm}
\caption{\label{cr} The spatial correlation function of dynamic
heterogeneity in real (top panel) and Fourier (bottom panel) spaces
reveals the growth of a dynamic length scale as $T \to 0$.  
Temperature decreases from left to right (top panel) and from bottom
to top (bottom panel). 
The dashed line in the bottom panel denotes 
the asymptotic behaviour $S(q,\tau,T)
\sim q^{-(2-\eta)}$, when $q \xi \gg 1$, with $\eta=-0.15$.}
\end{figure}

In Figure \ref{cr} we show the correlator $C(r,t,T)$ and the structure
factor $S(q,t,T)$ for different temperatures and fixed times
$t=\tau(T)$ where dynamic heterogeneity is maximal.  These
correlation functions clearly confirm the impression given by
Fig.~\ref{bubbles}, that a dynamic length scale associated with
spatial correlations of mobility develops and grows as $T$ decreases.
Note that at the lowest temperatures the structure factor does not
reach a plateau at low $q$. This because the system size we use,
although very large ($N=160^3$), is not sufficiently so to allow us to
probe the regime $q \xi \ll 1$. The necessary system sizes are simply
too large to simulate on such long time scales.

\begin{figure}
\psfig{file=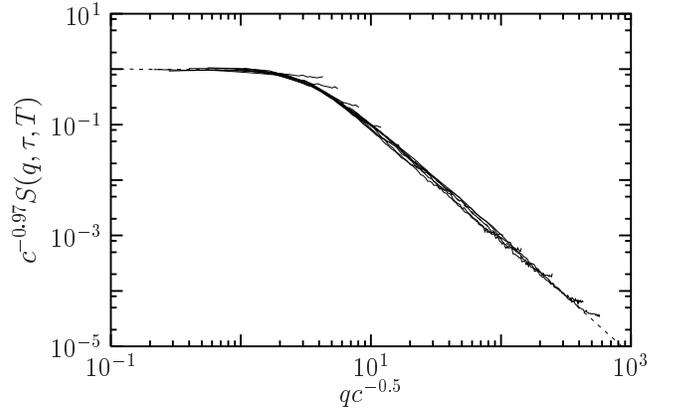,width=8.5cm}
\caption{\label{scal} Collapse of the dynamic structure factor
of Fig.~\ref{cr} using scaling laws (\ref{exp}) 
and taking $\gamma = 0.97$ and $\nu_\perp = 0.5$.
Dashed line is the scaling function $S(x)=1/(1+x^{2-\eta})$ with 
$\eta = -0.15$.}
\end{figure}

We can extract numerically the value of the dynamic length scale,
$\xi(T)$, at each temperature. To do so, we study in detail the shape
of the correlation functions shown in Fig.~\ref{cr}. As for standard
critical phenomena, we find that the dynamic structure factor can be
rescaled according to
\begin{equation}
S(q,t,T) \sim \chi(\tau,T) {\cal S} \left( q \xi \right),
\label{scalnum}
\end{equation}
where the scaling function $S(x)$ behaves as 
\begin{eqnarray}
S(x \to 0) \sim & const \\
S(x \to \infty) \sim & x^{2-\eta}.
\label{asympt}
\end{eqnarray}
Both the susceptibility $\chi$ and the dynamic length scale
$\xi$ estimated at time $t=\tau$ 
behave as power laws of the defect concentration,
\begin{equation}
\chi \sim c^{-\gamma}, \quad \quad \xi \sim c^{-\nu_\perp}.
\label{exp}
\end{equation}
These relations imply that the exponents $\gamma$ and $\nu_\perp$
should be numerically accessible by adjusting their values so that a
plot of $c^{\gamma} S$ versus $q c^{-\nu_\perp}$ is independent of
temperature. We show such a plot in Fig.~\ref{scal}, and we find that
the values $\gamma \approx 0.97$ and $\nu_\perp \approx 0.5$ lead to a
good collapse of the data.  The exponent $\gamma$ can be independently
and more directly estimated from Fig.~\ref{chi} by measuring the
height of the maximum of the susceptibility for various
concentrations. Fitting the result to a power law of $c$ gives $\gamma
\approx 0.96$, in reasonable agreement with the first value.  We find
also that the scaling function $S(x)$ is well-described by an
empirical form $S(x) = 1/(1+x^{2-\eta})$, consistent with
Eq.~(\ref{asympt}). Thus we can determine the value of the `anomalous'
exponent, $\eta$; we find $\eta \approx -0.15$.

As usual, it is difficult to estimate what constitutes the `best'
collapse of the data, and so determine accurately the errors in the
values of the exponents. Consequently, we are unable to determine
$\nu_\perp$ with sufficient accuracy to conclude that it agrees---or 
disagrees---with
the $d=3$ DP value, $\nu_\perp^{DP} \approx 0.58$.  It is also
difficult to compare $\gamma$ with its corresponding DP value, because
this would require one to know the anomalous exponent $\eta$
characterizing spatial correlations of the persistence function. From
a field theory perspective this is a formidable task. However, from
our numerics we have that $\eta \approx -0.15$, and so from scaling
arguments we find $\gamma = (2-\eta) \nu_\perp \approx 1.075$. This
estimates lies however 
on the `wrong' side of the classical value $\gamma^{cl}=1$ 
as compared to the numerical value obtained above.

We must conclude that numerical uncertainties
are too large, and deviations from classical behaviour too small to
make quantitative comparisons between DP and numerical exponents for
spatial correlations.  Plus, as we discussed above, our data may be
subject at very low temperature to small, but 
unknown, finite size effects.

We are nonetheless satisfied that the naive estimate $\nu_\perp = 1/d
= 1/3$~\cite{Ritort-Sollich}  that one gets by estimating the 
mean distance between defects
is invalidated by our numerical results.

\section{Conclusions}

We have derived a field theory for a kinetically constrained model
with isotropic facilitation, exemplified by the FA model. We have
studied the field theory via RG, and the lattice-based FA model via
numerical simulations. Our central results, briefly summarised in
Ref.~\cite{Whitelam-et-al}, are the following.

The RG treatment suggests that the low-$T$ dynamics is dominated by a
non-classical, zero-temperature critical point, which in turn implies
that correlation times, dynamic correlation lengths and
susceptibilities exhibit the scaling behaviour
\begin{equation}
\tau \sim c^{-\Delta}, \;\;\; \xi \sim c^{-\nu_{\perp}}, \;\;\; \chi
\sim c^{-\gamma} ,
\label{exps}
\end{equation}
with $\Delta = 1 + \nu_{\parallel}$. The Arrhenius behaviour of the
equilibrium concentration of excitations, $c \approx e^{-1/T}$, gives
rise to Arrhenius behaviour of the dynamics through (\ref{exps}). For
dimensions $d \geq 2$, the critical point is that of DP, while for
$d=1$ it is that of CDP. The upper critical dimension is $d_c=4$, so
that for dimensions $d \ge 4$ the exponents take classical values. For
$d=d_c=4$ the exponents are classical, augmented with the usual
logarithmic corrections~\cite{Amit}. For the time and space exponents
we have \cite{Hinrichsen}:
\begin{eqnarray}
\Delta &\approx& 3, \, 2.3, \, 2.1, \, 2 \;\;\;\;\; 
(d = 1, 2, 3, \ge 4), 
\label{ddelta} \\
\nu_{\perp} &\approx& 1, 0\, .73, \, 0.58, \,1/2.
\label{nnu}
\end{eqnarray}

We have also performed large-scale numerical simulations of the FA
model, which confirm many of the field-theoretic predictions. The
relaxation times of the FA model, Figs.\ \ref{tauT} and \ref{dpexp},
follow the scaling laws given by (\ref{exps}) and (\ref{ddelta}) in
all dimensions simulated ($d=1$ to $6$). The existence of an upper
critical dimension at $d_c=4$ is evident (see Fig.\ \ref{dpexp}). The
dynamics is increasingly heterogeneous and correlated in space as
temperature is decreased, as can be seen, for example, in pictures of
the local persistence (Fig.~\ref{bubbles}).  The structure factor for
this dynamic heterogeneity field in $d=3$ exhibits scale-invariance
(Fig.~\ref{scal}).

More extensive simulations are required in order to clarify two
further points. The spatial exponent obtained from the numerics is
$\nu_\perp \sim 0.5$, but we were unable to establish whether this
number agrees precisely with the $d=3$ DP value of
$\nu_\perp=0.58$. We also caution the reader that there may exist,
even for $d>2$, a crossover from early-time CDP behaviour to
intermediate-time DP behaviour, as is the case for some systems in,
ostensibly, the DP universality class~\cite{Hinrichsen}.

Our work shows that standard theoretical methods, such as the
renormalization group, can be used to analyze coarse-grained models of
glass-forming supercooled liquids~\cite{fss,steve}. It supports the
view that the dynamics of glass-formers is in many respects similar to
that of standard critical phenomena, such as reaction-diffusion
systems~\cite{Tauber-review,Lexie}. We have found, numerically and
analytically, that the FA model and its associated field theory
possess a zero-temperature critical point, in agreement with results
obtained by other means~\cite{Ritort-Sollich}. Rigorous results
confirm the existence of a $T=0$ critical point in other kinetically
contrained systems, such as the East model~\cite{Aldous-Diaconis}, and
an analogous maximal-density critical point in the Kob-Andersen model
\cite{Toninelli-et-al}. Extending the field theory treatment to models
of fragile glass-forming liquids, such as the East model \cite{Jackle}
and its generalizations \cite{Garrahan-Chandler-2}, constitutes an
interesting challenge.

Finally, our results provide some insight into the physical meaning of
fragility, in the Angell sense~\cite{Angell}.  First, we have shown
here and
elsewehere~\cite{Garrahan-Chandler,Garrahan-Chandler-2,Berthier-Garrahan}
that strong systems show fluctuation-dominated heterogeneous dynamics,
in a similar manner to fragile systems. This contradicts the popular
view that `cooperativity', `fragility', and `heterogeneity' are
different facets of the same concept~\cite{Ediger-et-al,Angell}.

In our view, the difference between strong and fragile liquids is in the strength of fluctuation
effects. For example, the breakdown of the Stokes-Einstein relation
observed in fragile liquids~\cite{Chang-Sillescu,Swallen-et-al} should
also be observed in strong ones \cite{Jung-et-al}, but the effect will
be less striking.  However, since strong systems such as the FA model
are characterized by a constant dynamic exponent, we expect that
typical length scales at the glass transition are typically larger in
strong glass-formers than in fragile ones. Detailed studies of dynamic
heterogeneity in atomistic models of strong liquids should be able to
test these predictions~\cite{Vogel-Glotzer}, while experimental
investigations of strong glass-formers would also be very welcome.

\acknowledgments

We thank G.~Biroli, J.-P.~Bouchaud, P. Calabrese, J.L.~Cardy,
D.~Chandler, M.~Kardar, W.~Kob, B.~Ruffl\'e and O.~Zaboronski for discussions and comments. We acknowledge financial support from EPSRC Grants No.\
GR/R83712/01 and GR/S54074/01, Marie Curie Grant No.\
HPMF-CT-2002-01927 (EU), CNRS France, Linacre and Worcester Colleges
Oxford, University of Nottingham Grant No.\ FEF 3024, and numerical
support from the Oxford Supercomputing Centre.

\end{document}